\documentclass[]{mn2e}

\usepackage{graphicx}
\usepackage{latexsym}
\usepackage{float}

\title[The Magnetic Field of the H~II Region NGC 6334A]{The Magnetic Field of the H~II Region NGC 6334A
from Faraday Rotation}
\author[L. F. Rodr\'\i guez et al.]{L. F. Rodr\'\i guez$^{1,2}$\thanks{E-mail:
l.rodriguez@crya.unam.mx (LFR)}, Y. G\'omez$^{1}$ and D.
Tafoya$^{3}$ \\
$^{1}$Centro de Radiostronom\'{\i}a
y Astrof\'\i sica, 
Universidad Nacional Aut\'onoma de M\'exico, Apdo. Postal 3-72, Morelia, Michoac\'an \\ 58089, M\'exico \\
$^{2}$Astronomy Department, Faculty of Science, King Abdulaziz University, 
P.O. Box 80203, Jeddah 21589, Saudi Arabia \\
$^{3}$Department of Physics and Astronomy, Graduate School of Science and Engineering,  
Kagoshima University, 1-21-35 Korimoto, \\ Kagoshima 890-0065, Japan}

\begin{document}

\date{Accepted 2011 December 15. Received 2011 December 14; in original form 2011 October 11}

\pagerange{\pageref{firstpage}--\pageref{lastpage}} \pubyear{2011}

\maketitle

\label{firstpage}

\begin{abstract}
We have studied the polarization characteristics and Faraday
rotation of the extragalactic radio source  
J17204$-$3554, that appears projected on the north lobe of the galactic H~II region
NGC~6334A. From observations made with the Very
Large Array at 6.0 and 3.6 cm in three different epochs (1994, 1997, and 2006), we estimate
a rotation measure of +5100$\pm$900 rad m$^{-2}$ for the extragalactic source.
This large rotation measure implies a line-of-sight average magnetic field of
$B_\parallel \simeq +36\pm6$ $\mu$G, the largest obtained by this method
for an H~II region. NGC~6334A is significantly denser than other H~II regions
studied and this larger magnetic field is expected on the grounds of
magnetic flux conservation. The ratio of thermal to magnetic pressure is
$\sim$5, in the range of values determined for more diffuse H~II regions. 
%With the new generation of ultrasensitive
%radio interferometers it will become feasible to detect the Faraday rotation
%of fainter extragalactic sources projected behind denser compact H~II regions
%to improve our understanding of the relation between electron densities and magnetic fields.

\end{abstract}


\section{Introduction}

The magnetic field of astronomical sources is an important but in general difficult
parameter to measure. In their pioneering work,
Heiles \& Chu (1980) and Heiles, Chu, \& Troland (1981) used the Faraday
rotation of extragalactic background sources behind four diffuse H~II
regions (S232, S117, S119, and S264) to determine rotation
measures with absolute values in the order of $10^2 - 10^3$ rad m$^{-2}$. These rotation measures
were then used to estimate the strength of the ordered magnetic field along the line of sight,
$B_\parallel$,
to have values in the range of $|B_\parallel| \simeq$ 1 to 20 $\mu$G for regions with electron densities
in the range of 2 to 14 $cm^{-3}$. As part of their
polarization survey of the Galactic plane,
Sun et al. (2007) also used Faraday rotation to
estimate the strength of the ordered magnetic field 
along the line of sight to two diffuse H~II regions, G124.9+0.1 and G125.6-1.8.
In the case of G124.9+0.1 (with an estimated electron density
of 1.6 $cm^{-3}$), this magnetic field was
found to be $+3.9~ \mu$G, while in the case of G125.6-1.8 (with an upper
limit to the electron density of 0.84 $cm^{-3}$) a lower
limit of $+6.4~ \mu$G was obtained. More recently, Harvey-Smith et al. (2011) studied
the line-of-sight magnetic fields in five large-diameter Galactic H~II regions
(Sh 2-27, Sh 2-264, Sivan 3, Sh 2-171, and Sh 2-220), with electron densities in
the range of 0.8 $-$ 30 $cm^{-3}$.
Using the Faraday rotation technique, they find $|B_\parallel|$ values in the
range from 2 to 6 $\mu$G, which are similar to the typical magnetic field 
strength in the diffuse interstellar medium.
The Faraday rotation method has not been used for more compact
and denser H~II regions since the probability of having a bright, linearly
polarized background source behind a small solid angle is low.

In this paper we present determinations of the rotation measure 
at different epochs and frequencies 
in the direction of the AGN J17204-3554 (Bykov et al 2006; Krivonos et 
al 2007; Feigelson et al. 2009). This extragalactic object ($\alpha(2000) = 17^h~ 20^m~ 21\rlap.{^s}81,
\delta(2000) = -35^\circ ~ 52{'} 48\rlap.{''}2$; 
$l = 351\rlap.^\circ277, b = +00\rlap.^\circ678$) is a bright radio source 
(Rodr\'\i guez et al. 1980) and was originally believed to be embedded
in the molecular cloud that contains the NGC~6334 galactic star-forming complex.
It is located in the direction of the northern lobe of the bipolar H~II region
NGC 6334B (Moran et al. 1990) and it is known to be heavily scattered by the
intervening plasma (Trotter et al. 1998). 

Moran et al. (1990)
estimate an emission measure of $6 \times 10^4~cm^{-6}~pc$
for the H~II region NGC~6334A along the line of sight to J17204-3554.
Assuming that the depth of the ionized volume is comparable to the width of
NGC~6334A in the plane of the sky (about 1$'$ or 0.5 pc at a distance of
1.7 kpc) and a homogeneous medium, we estimate an electron density of
$\sim$350 $cm^{-3}$ for the intervening H~II region.    
Since this electron density is much larger than the values in the lines of sight
sampled by Heiles \& Chu (1980), Heiles, Chu, \& Troland (1981), Sun et al. (2007), and
Harvey-Smith et al. (2011)
we expected a larger magnetic field strength determination from the Faraday
rotation measurements.

\begin{table*}
\centering
 \begin{minipage}{140mm}
       \caption{VLA Archive Observations of J17204$-$3554 Used in this Paper}
       \begin{tabular}{lccccccc}
       \hline
       & VLA &     & Wavelength &  Amplitude & Phase  & Parallactic Angle & Time On-source \\
 Epoch & Configuration & Project  & (cm) & Calibrator & Calibrator & Range ($^\circ$) & (minutes) \\ 
\hline
1994 Apr 28 & A & AM447 & 6.0 & 1331+305 & 1744$-$312 & 23 &  50  \\
1997 Feb 02 & BnA & AT202 & 6.0 & 1331+305 & 1733$-$130 & 43 & 39 \\ 
1997 Feb 02 & BnA & AT202 & 3.6 & 1331+305 & 1733$-$130 & 47 & 60 \\ 
1997 Feb 02 & BnA & AT202 & 2.0 & 1331+305 & 1733$-$130 & 61 & 100 \\ 
2006 Aug 17 & B   & S7810 & 3.6 & 0137+331 & 1832$-$105\footnote{For these
observations 1832$-$105 was used to correct for instrumental polarization
and J17204$-$3554, now included in the list of VLA calibrators, was self-calibrated.} & 63 & 25 \\
\hline
\end{tabular}
\end{minipage}
\end{table*}

\section{VLA Archive Observations}

We searched in the archive of the Very Large Array (VLA) for continuum
data sets with the following
characteristics. They should have been obtained in the high-angular resolution A or B
configurations to minimize the effect of the strong extended emission in the region.
The observations were also required to have been taken over several hours and to
include a phase calibrator observed over a wide range of parallactic angles, to
correct for instrumental polarization. We found three adequate data sets,
whose parameters are summarized in Table 1. The data were edited and
calibrated in amplitude, phase and polarization using the Astronomical
Image Processing System (AIPS) package of the US National Radio
Astronomy Observatory (NRAO), following the standard VLA procedures.
The polarization calibration was performed using the observations of the amplitude
calibrator to determine the absolute polarization angle, while the observations of the phase
calibrator were used to determine and correct the antenna-based leakage terms that produce
instrumental polarization. 

We determined the Stokes parameters I, Q, and U of the
sources observed by fitting directly the images with Gaussian
functions using the task JMFIT of AIPS. The errors in the Stokes
parameters are estimated from the intensity and angular dimensions of the sources
and from the noise in the images
following Condon (1997). These parameters,
as well as the derived percentages and position angles
of the linear polarization are given in Tables 2 to 4.
The errors in the percentages and the position angles
come from the propagation of the errors on the Stokes parameters. 
We list separately these parameters for the two IFs available at the VLA
(separated by 50 MHz) at each band.

\begin{table*}
\centering
 \begin{minipage}{140mm}
 \caption{Linear Polarization Parameters of 1744-312 and J17204-3554 on 1994 April 28}
  \begin{tabular}{lcccccc}
     \hline
Source & Frequency (GHz)  & I (Jy)\footnote{Stokes parameters I, Q, and U.}
& Q (mJy)$^a$ & U (mJy)$^a$& Polarization (\%)\footnote{Percentage of linear polarization.}
& P.A. ($^\circ$) \footnote{Position angle of the linear polarization.} \\
\hline
1744$-$312 & 4.8351 & 0.386$\pm$0.001 & $-$5.4$\pm$0.2 & $-$1.3$\pm$0.2 & 
1.44$\pm$0.05 & $-$83.2$\pm$2.1 \\
1744$-$312 & 4.8851 & 0.387$\pm$0.001 & +5.7$\pm$0.2 & +0.0$\pm$0.1 &
1.47$\pm$0.05 &  $-$90.0$\pm$1.5 \\

\hline 
J17204$-$3554 & 4.8351 & 0.470$\pm$0.002 & $-$0.1$\pm$0.1 & $-$2.8$\pm$0.1 &
0.60$\pm$0.02 &  $-$46.0$\pm$2.0 \\
J17204$-$3554 & 4.8851 & 0.473$\pm$0.002 & $-$2.4$\pm$0.1 &  $-$2.1$\pm$0.1 &
0.67$\pm$0.02 &  $-$69.4$\pm$1.8 \\

\hline
\end{tabular}
\end{minipage}
\end{table*}

\begin{table*} 
\centering
 \begin{minipage}{140mm}
      \caption{Linear Polarization Parameters of 1733-130 and J17204-3554 on 1997 February 02}
    \begin{tabular}{lcccccc}
    \hline
   Source & Frequency (GHz)  & I (Jy)\footnote{Stokes parameters I, Q, and U.}
   & Q (mJy)$^a$ & U (mJy)$^a$& Polarization (\%)\footnote{Percentage of linear polarization.} 
   & P.A. ($^\circ$) \footnote{Position angle of the linear polarization.} \\
    \hline
    1733$-$130 &  4.8351 & 9.274$\pm$0.007 & $-$72.7$\pm$1.0 &
    +128.6$\pm$1.9 & 1.59$\pm$0.02 & +59.7$\pm$0.5 \\
    1733$-$130 &  4.8851 & 9.334$\pm$0.006 & $-$72.5$\pm$1.2 &
    +120.0$\pm$0.6 & 1.50$\pm$0.01 & +60.6$\pm$0.4 \\ 
    1733$-$130 & 8.4351 & 11.804$\pm$0.005 & $-$183.4$\pm$0.4 &
    +104.6$\pm$0.3 & 1.79$\pm$0.01 & +75.2$\pm$0.1 \\
    1733$-$130 & 8.4851 & 11.830$\pm$0.010 & $-$182.2$\pm$0.9 &
    +107.3$\pm$0.9 & 1.79$\pm$0.01 & +74.8$\pm$0.2 \\
    1733$-$130 & 14.9149 & 13.170$\pm$0.009 & $-$232.0$\pm$0.8 &
    +167.5$\pm$0.7 & 2.17$\pm$0.01 & +72.1$\pm$0.1 \\
    1733$-$130 & 14.9549 & 13.154$\pm$0.008 & $-$234.1$\pm$1.2 & 
    +169.6$\pm$1.6 & 2.20$\pm$0.01 & +72.0$\pm$0.3 \\
    \hline
    J17204$-$3554 &  4.8351 & 0.715$\pm$0.007 & +2.1$\pm$0.1 &
    $-$2.5$\pm$0.1 & 0.46$\pm$0.02 & $-$25.0$\pm$1.8 \\
    J17204$-$3554 &  4.8851 & 0.729$\pm$0.007 & +0.2$\pm$0.1 &
    $-$3.6$\pm$0.1 & 0.49$\pm$0.02 & $-$43.4$\pm$1.6 \\
    J17204$-$3554 & 8.4351 & 0.671$\pm$0.006 & $-$11.1$\pm$0.1 &
    +6.3$\pm$0.1 & 1.90$\pm$0.02 & +75.2$\pm$0.4 \\
    J17204$-$3554 & 8.4851 & 0.669$\pm$0.005 & $-$9.9$\pm$0.1 &
    +8.4$\pm$0.1 & 1.94$\pm$0.02 & +69.8$\pm$0.4 \\
    J17204$-$3554 & 14.9149 & 0.552$\pm$0.006 & +11.9$\pm$0.2 &
    $-$7.6$\pm$0.2 & 2.56$\pm$0.04 & $-$16.3$\pm$0.8 \\
    J17204$-$3554 & 14.9549 & 0.554$\pm$0.006 & +11.7$\pm$0.2 &
    $-$8.0$\pm$0.2 & 2.56$\pm$0.04 & $-$17.2$\pm$0.8 \\

	    \hline
\end{tabular}
\end{minipage}
\end{table*}

\begin{table*}
\centering
 \begin{minipage}{140mm}
  \caption{Linear Polarization Parameters of 1744-312 and J17204-3554 on 2006 August 17}
    \begin{tabular}{lcccccc}
	 \hline
 Source & Frequency (GHz)  & I (Jy)\footnote{Stokes parameters I, Q, and U.}
 & Q (mJy)$^a$ & U (mJy)$^a$& Polarization (\%)\footnote{Percentage of linear polarization.}
 & P.A. ($^\circ$) \footnote{Position angle of the linear polarization.} \\
 \hline
 1744$-$312 & 8.4351 & 0.594$\pm$0.001 & +1.6$\pm$0.2 & $-$3.7$\pm$0.2 &
 0.68$\pm$0.03 & $-$33.3$\pm$2.8 \\
 1744$-$312 & 8.4851 & 0.592$\pm$0.001 & +1.5$\pm$0.2 & $-$4.2$\pm$0.2 &
 0.75$\pm$0.03 & $-$35.2$\pm$2.6 \\

 \hline 
 J17204$-$3554 & 8.4351 & 0.488$\pm$0.004 & $-$7.6$\pm$0.1 & +0.0$\pm$0.1 & 
 1.56$\pm$0.02 & +90.0$\pm$0.8 \\
 J17204$-$3554 & 8.4851 & 0.484$\pm$0.005 & $-$8.1$\pm$0.1 & +1.6$\pm$0.1 & 
 1.71$\pm$0.02 & +84.4$\pm$0.7 \\

 \hline
 \end{tabular}
 \end{minipage}
 \end{table*}

We searched for significantly different position angles in the two IFs of the data at
each observed wavelength and epoch. Only in the case of the source of
interest J17204$-$3554 (epochs 1994 and 1997 at
6.0 cm and epoch 1997 and 2006 at 3.6 cm) and the phase
calibrator 1744$-$312 (epoch 1994 at 6.0 cm) 
we find significant differences. 
We determined the rotation measures (RM) from these observations using

\begin{equation}
RM = -{{\Delta \theta~ \nu} \over {2~ \lambda^2~ \Delta \nu}}~rad~m^{-2},
\end{equation}
 where
 $\Delta \theta$ is the difference in position angles given in radians,
 $\nu$             is the central frequency in MHz,
 $\lambda$    is the central wavelength in meters, and
 $\Delta \nu$        is the frequency difference in MHz (that in general, can be given in the same
 units used for $\nu$).
With this definition, positive and negative RMs indicate average
magnetic fields pointed towards and away from the observer, respectively. 

For the 1994 observations of 1744$-$312 at 6 cm, we derive RM = +1500$\pm$600 rad m$^{-2}$,
in agreement with the much more accurate value of +1883$\pm$3 rad m$^{-2}$
derived by Roy et al. (2005) from observations made from 4.8 to 8.5 GHz.
Our technique is insensitive to small rotation measures and the upper limits 
for the rotation measure derived from the 1997 observations
of 1733$-$130 at 6.0 cm, $|$RM$|$ $\leq$1000 rad m$^{-2}$ (3-$\sigma$ upper limit) are
also consistent with the value of $\sim$$-$60 rad m$^{-2}$ reported in the
rotation measure catalog of Taylor et al. (2009) and previously by Rusk (1988).

\begin{table}
\centering
 \begin{minipage}{140mm}
\caption{Rotation Measures derived for J17204$-$3554}
      \begin{tabular}{lcc}
    \hline
    &   Wavelength &  Rotation \\
   Epoch  & (cm) & Measure(rad m$^{-2}$) \\
     \hline
     1994 Apr 28 & 6.0 & +5200$\pm$600  \\
     1997 Feb 02 & 6.0 & +4100$\pm$500  \\
     1997 Feb 02 & 3.6 & +6300$\pm$700 \\
     2006 Aug 17 & 3.6 & +6600$\pm$1200 \\
      \hline
      \end{tabular}
      \end{minipage}
      \end{table}

In Table 5 we report our main result: the rotation measures derived for J17204$-$3554.
We note that the two 6.0 cm determinations are consistent among them
and the same happens for the two 3.6 cm determinations. However, the 3.6 cm
determinations appear to be $\sim$40\% larger than those at
6.0 cm and this effect deserves further attention. Law et al. (2011) have compared
observations made in the 1 to 2 GHz range with the Allen Telescope Array
and in the 5 to 22 GHz range with the Very Long Baseline Array and
conclude that, as observed for J17204$-$3554, the rotation measure values and 
the fractional polarization are generally larger at higher frequencies.
In the case of J17204$-$3554 this effect could be related to the fact that
plasma scattering makes the source larger at the lower frequencies (Trotter et al.
1998). We may then be averaging over a larger volume 
at the lower frequencies and obtaining a lower value for the line-of-sight
magnetic field. In any case, given the
modest signal-to-noise ratio of our results, we will adopt
as the rotation measure for J17204$-$3554 the value
of +5100$\pm$900 rad m$^{-2}$, the weighted average of the four determinations.

The absolute position angles of the linear polarization for J17204$-$3554
at a given frequency change significanty from one epoch to the other (compare,
for example, the position angles at 4.8351 or 4.8551 GHz 
for 1994 April 28 and 1997 February 02).
It is known that the absolute position angles of quasars
at a given frequency can change by as much
as several tens of degrees
over timescales of months (see Zavala \& Taylor 2001
and also The VLA/VLBA Polarization Calibration Page
in http://www.aoc.nrao.edu/~smyers/calibration/).
In the case of a source with a large rotation measure such as J17204$-$3554,
these variations most probably come from small changes in the parameters
of the Faraday screen with time. For a rotation measure of +5100 rad m$^{-2}$, a variation
of only $\sim$2\% in its value will produce a change of $\sim 20^\circ$
in the position angle at 4.8 GHz. 

\section{Interpretation}

The rotation measure estimated for J17204$-$3554 is quite large. 
The largest rotation measures known are usually found
for extragalactic sources with a similar location to that of
J17204$-$3554, that is, in the galactic
plane toward the centre of the Galaxy. 
Brown et al. (2007) presented Faraday rotation measures for 148 extragalactic 
radio sources behind the southern Galactic plane ($253^\circ \leq l \leq 356^\circ$; 
$|b|\leq 1.5^\circ$). This region includes J17204$-$3554 but this source
is not measured by Brown et al. (2007), possibly because of the complex emission 
from the NGC~6334A H~II region. Of these 148 sources, only two have absolute rotation measures in excess
of 1000 rad m$^{-2}$ (+1113$\pm$11 rad m$^{-2}$ for G296.90+0.14 and $-$1035$\pm$16 rad m$^{-2}$
for G343.29+0.60). In the surroundings of J17204$-$3554 the typical absolute value of the
rotation measure is $\sim$300 rad m$^{-2}$. We can then propose that most of the rotation
measure seen toward J17204$-$3554 is produced by the H~II region NGC~6334A.
Roy et al. (2005; 2008)  
determined the rotation measure toward 60 background extragalactic components in 
the region defined by $-6^\circ \leq l \leq 6^\circ$ , $-2^\circ \leq b \leq 2^\circ$, 
that surrounds the
galactic centre, where the largest rotation measures are expected. In this extreme sample,
there is only one source with a rotation measure comparable with J17204$-$3554 (the
source G358.917+0.072,
about one degree away from Sgr A*, with a rotation measure of $\sim$+4800 rad m$^{-2}$).

If we then assume that practically all the Faraday rotation observed toward J17204$-$3554 is produced by
the H~II region NGC~6334A we can estimate the line-of-sight magnetic field of this last
source. The rotation measure is given by

\begin{equation}
RM = 0.81 \int n_e~B_\parallel~dl~rad~m^{-2}
\end{equation}
 where
  $n_e$ is the electron density in the medium producing the
  Faraday rotation, given in cm$^{-3}$,
  $B_\parallel$  is the parallel component of the magnetic field in $\mu$G, and
  $dl$    is the differential of path length along the line of sight, given in pc.

Assuming a homogeneous medium and the values estimated by Moran et al. (1990) of
$n_e \simeq 350$ cm$^{-3}$ and $l$ = 0.5 pc, we obtain $B_\parallel \simeq +36\pm6$ $\mu$G,
with this average magnetic field pointing toward us.
This is the largest value of $B_\parallel$ obtained from Faraday rotation across an H~II
region. Harvey-Smith et al. (2011) present a compilation of values determined with this technique
and find them to be typically located in the ranges of $0.6~\mu G \leq B_\parallel \leq 10~\mu G$ and
$1~cm^{-3} \leq n_e \leq 20~cm^{-3}$. 
Harvey-Smith et al. (2011) have combined the H~II Faraday rotation data with HI Zeeman effect observations
to discuss line-of-sight magnetic fields 
for densities (electron or atomic) in the range of $1~cm^{-3} \leq n \leq 200~cm^{-3}$.
They conclude that the slope of magnetic field versus density in this low-density regime is quite flat.
A value of $|B_\parallel| \simeq 3\pm3~\mu G$ could fit most of the available data.
However, the value of $B_\parallel$ for 
NGC~6334A is substantially
larger and indicates that it is very important to extend
the Faraday rotation technique to denser H~II regions, with $n_e > 100~cm^{-3}$.

Unfortunately, denser H~II regions are typically smaller and this makes the probability of
finding a bright polarized 
background source in their line of sight practically negligible. Fortunately, with the advent
of ultrasensitive radio interferometers such as the EVLA and eMERLIN, one will soon be able to
detect and study much weaker background sources, extending the Faraday rotation technique to 
higher electron densities.

Finally, following Harvey-Smith et al. (2011), we estimate the
ratio of thermal to magnetic pressure, $\beta_{th}$, for NGC~6334A.
This ratio is given by

\begin{equation}
\beta_{th} = {{16 \pi~n_e~k~T_e} \over {B^2}}
\end{equation}
 where
  $k$ is Boltzmann's constant,
   $T_e$ is the electron temperature, assumed to be $10^4$ K, and
   $B$ is the total magnetic field, assumed to be equal to $\sqrt{3} B_\parallel$.

From the values given in the paper, we obtain $\beta_{th} \simeq$ 5,
in the range of $2 \leq \beta_{th} \leq 22$ obtained by 
Harvey-Smith et al. (2011) for the five large-diameter Galactic H~II regions
studied by them.

\section{Conclusions}

Our main results can be summarized as follows.

\begin{enumerate}
  \item We studied the Faraday
  rotation of the extragalactic radio source  
  J17204-3554, that appears projected on the north lobe of the galactic H~II region
  NGC~6334A, estimating a 
  rotation measure of +5100$\pm$900 rad m$^{-2}$. This is one of the largest
  rotation measures found for any type of source.
  \item This rotation measure implies a line-of-sight magnetic field of
  $B_\parallel \simeq +36\pm6$ $\mu$G for NGC~6334A, the largest obtained by this method
  for an H~II region.
  \item For the ratio of thermal to magnetic pressure we obtain $\beta_{th} \simeq$ 5,
  in the range of $2 \leq \beta_{th} \leq 22$ obtained by 
  Harvey-Smith et al. (2011) from studies of five large-diameter Galactic H~II regions.
  \item These results suggest that it is necessary to study
  much weaker background sources behind denser H~II regions, extending the Faraday rotation technique to 
  higher electron densities. This possibility will become feasible with
  the ultrasensitive arrays like eMERLIN and the EVLA.

\end{enumerate}

\section*{Acknowledgments}

LFR acknowledges the financial support of DGAPA, UNAM and CONACyT, 
M\'exico.

\bsp

\label{lastpage}

\end{document}